\documentclass[aps,prl,twocolumn,showpacs,floatfix,preprintnumbers,amsmath,
amssymb,superscriptaddress]{revtex4}
\usepackage{graphicx}
\usepackage{color}

\newcommand{\beq}{\begin{equation}}
\newcommand{\eeq}{\end{equation}}
\newcommand{\beqn}{\begin{eqnarray}}
\newcommand{\eeqn}{\end{eqnarray}}
\newcommand{\bearr}{\begin{array}}
\newcommand{\enarr}{\end{array}}

\begin{document} 

\title{Negative Temperature States in the Discrete Nonlinear Schr\"odinger
Equation}
      
\author{S. Iubini}
\affiliation{Dipartimento di Fisica e Astronomia -- CSDC, Universit\`a di Firenze and
INFN Sezione di Firenze, via Sansone 1, I-50019 Sesto Fiorentino, Italy}
\affiliation{CNR - Istituto dei Sistemi Complessi, via Madonna del Piano 10, 
I-50019 Sesto Fiorentino, Italy}
\author{R. Franzosi}
\affiliation{CNR - Istituto dei Sistemi Complessi, via Madonna del Piano 10, 
I-50019 Sesto Fiorentino, Italy}
\author{R. Livi}
\affiliation{Dipartimento di Fisica e Astronomia -- CSDC, Universit\`a di Firenze and
INFN Sezione di Firenze, via Sansone 1, I-50019 Sesto Fiorentino, Italy}
\author{G.-L. Oppo}
\affiliation{ICS, SUPA and Department of Physics, 
University of Strathclyde, 107 Rottenrow, Glasgow G4 ONG, U.K.}
\author{A. Politi}
\affiliation{ICSMB, SUPA and Department of Physics, University of Aberdeen AB24 3UE,
U.K.}
\affiliation{CNR - Istituto dei Sistemi Complessi, via Madonna del Piano 10, 
I-50019 Sesto Fiorentino, Italy}

\begin{abstract}

We explore the statistical behavior of the discrete nonlinear Schr\"odinger equation.
We find a parameter region where the system evolves towards a state characterized by 
a finite density of breathers and a negative temperature.
Such a state is metastable but the convergence to equilibrium occurs on astronomical 
time scales and becomes increasingly slower as a result of a coarsening processes. 
Stationary negative-temperature states can be experimentally generated via boundary 
dissipation or from free expansions of wave packets initially at positive temperature 
equilibrium.
\end{abstract}

\maketitle 
The discrete nonlinear Schr\"odinger (DNLS) equation has been widely
investigated as a semiclassical model for Bose-Einstein condensates (BEC) in
optical lattices and for light propagation in arrays of optical waveguides
\cite{kevre}. It exhibits peculiar solutions in the form of lattice solitons,
also known as breathers, or intrinsically localized nonlinear excitations
\cite{NonLin11}. In the first statistical-mechanics study of the DNLS equation, 
Rasmussen {\it et al.} \cite{Cretegny} identified a region $\mathrm{R_n}$ in the 
parameter space that was conjectured to be characterized by negative temperature 
(NT) states. NT states have attracted the curiosity of researchers since pioneering
works in systems of quantum nuclear spins \cite{ppr}. In a series of recent
papers, Rumpf argued that there are no NT equilibrium states in $\mathrm{R_n}$.
In fact, he showed that the maximum entropy (i.e. equilibrium) state is
characterized by an infinite-temperature background superposed to a single
breather that collects the ``excess" energy \cite{rumpf}. Convergence 
to such equilibrium states, however, has never been observed, since the transient 
may last for astronomical times \cite{rasm1}. Therefore it is
legitimate to ask what happens over physically accessible time scales. 
The occurrence of long transients is not rare in statistical mechanics:
it may be due to coarsening, nucleation, presence of free-energy
barriers, or the stability of some modes. Within the physical setups
that are closest to the DNLS, a nonexponential relaxation was first found in
chains of nonlinear oscillators and shown to originate from the presence of 
long-lived localized solutions \cite{tsironis}. Slow relaxations have been also
found in Heisenberg spin chains and traced back to the existence of two 
conservation laws \cite{RumpfN}. This latter example is particularly instructive,
since it is the context where theoretical arguments, later applied to the
DNLS, have been developed. The approach is well suited for the identification
of the equilibrium state by maximizing the entropy given the constraints
imposed by the conservation laws. It is however of little help to investigate
the convergence properties to equilibrium.
In this respect, the implementation of a microcanonical Monte Carlo (MMC) method
turns out to be quite useful, in that it allows uncovering (in the so-called
weak-coupling limit) a first source of ``slowness" in an underlying
coarsening phenomenon that follows from the subdiffusive behavior of the
breather amplitude. As a result, we find that the density of breathers vanishes
in time as $t^{-\alpha}$ with an exponent $\alpha\approx0.37$ that is
reminiscent of the exponent $1/3$ of the Cahn-Hilliard model \cite{Bray}. 
Note, however, that dynamical mechanisms which strongly inhibit the diffusion
of the breather amplitude are even more effective. With the help of an explicit 
expression for the microcanonical temperature \cite{franzo}, we are able to show that
a broad class of initial conditions (IC) of the DNLS equation converges towards a 
well defined thermodynamic state characterized by a negative temperature and a
finite density of breathers. There is no reason to dismiss the theoretical
arguments of \cite{rumpf}, although it appears that the dynamical freezing of the
high-amplitude breathers slows down the evolution so much to make the
convergence to equilibrium unobservable. Altogether this phenomenon
is reminiscent of aging in glasses, but a more detailed analysis is
required to frame the analogy on more firm grounds.
Finally, we briefly comment on a simple strategy to generate experimentally
metastable NT states in BEC in optical lattices and in arrays of optical
waveguides. Our scheme is far simpler than the thermalization methods recently
proposed for BEC trapped in lattices which require non-trivial Mott states
followed by sweeps across Feshbach resonances to invert the sign of the atomic
interaction \cite{mosk05}. 

\begin{figure}
\includegraphics[clip, width=7cm]{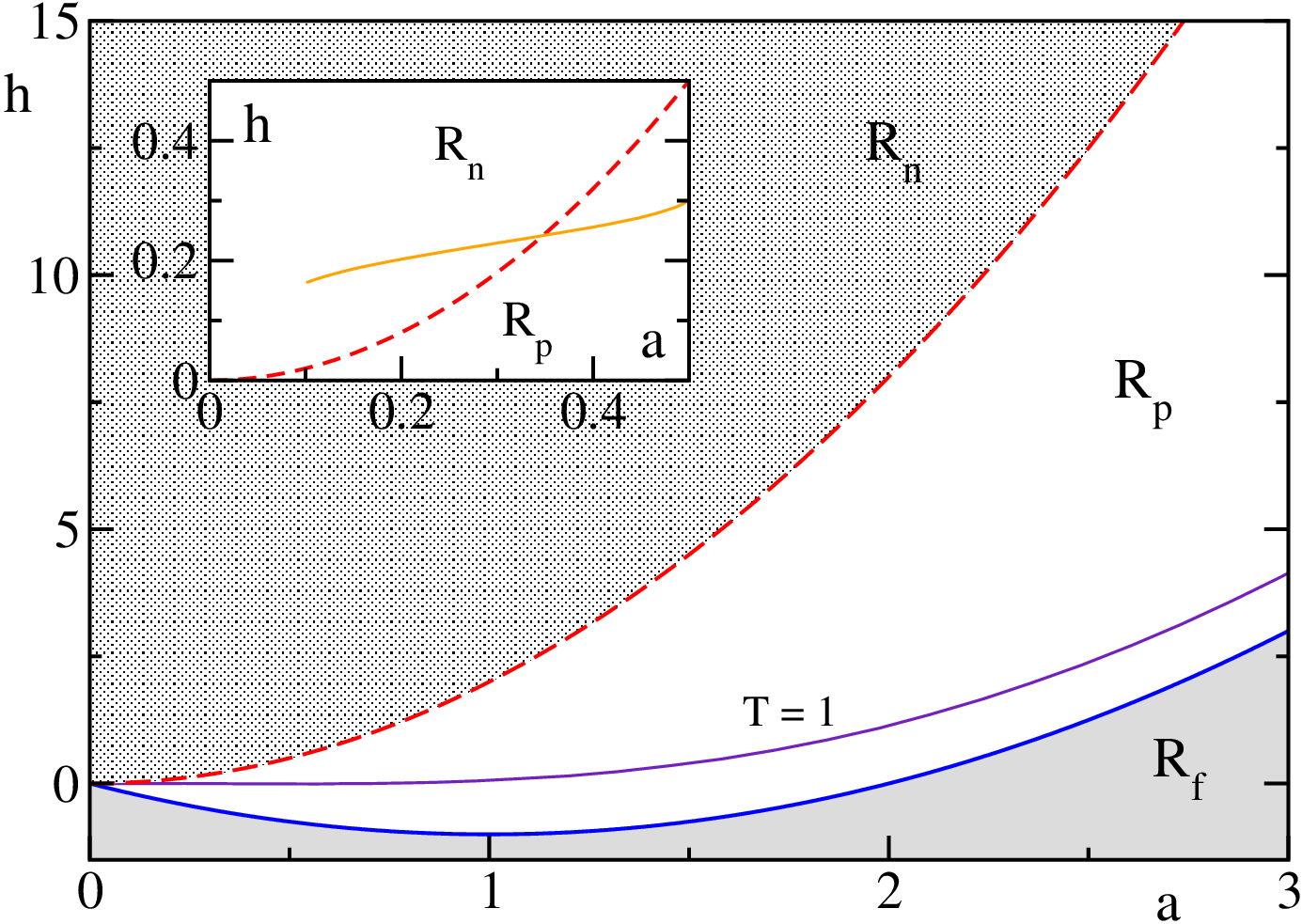}
\caption{(Color online) Phase diagram ($a$,$h$) \cite{Cretegny}. 
We show the isothermal lines $T=0$ (solid blue), $T=1$ (solid purple) and
$T=\infty$ (dashed red). In the inset: average trajectory (over an
ensemble of 500 realizations) of a chain with $N=200$ in the presence of
boundary dissipation. 
}
\label{paramsp}
\end{figure}

{\it Model and state of the art.} With reference to the standard canonical
coordinates, the Hamiltonian of a DNLS chain writes
\beq 
H(p,q) = \sum_{j}  (p_j^2+q_j^2)^2 +
2 \sum_{j}(p_jp_{j_{+1}}+q_jq_{j_{+1}}) \, ,
\label{Hdnlse}
\eeq
where the index $j$ numbers the sites from 1 to $N$ and periodic boundary
conditions are assumed. The sign of the first term in the r.h.s. is
assumed to be positive, since we refer to the case of a repulsive-atom BEC,
while the sign of the hopping term is irrelevant, due to the symmetry
associated with the canonical (gauge) transformation
$\phi_j\to \phi_j + \pi$ (where $p_j+iq_j\equiv u_j {\rm e}^{i\phi j}$
represents the wave function at site $j$). The model
possesses two conserved quantities: the energy $H$ and the number of particles
(or, wave action)
$A = \sum_j u_j^2$. The thermodynamic properties of this model are summarized 
in a phase-diagram of energy ($h$) versus particle ($a$) density 
per site \cite{Cretegny}. In Fig.~\ref{paramsp}, one can identify three regions: 
(i) a forbidden region $\mathrm{R_f}$, that lies below the ground state;
(ii) an intermediate positive temperature $\mathrm{R_p}$ region; (iii)
the above mentioned region $\mathrm{R_n}$, that is investigated in this Letter.
The presence of low relaxation phenomena in $\mathrm{R_n}$ was first noticed in
\cite{Cretegny} by monitoring the probability density of the local amplitudes
$A_j \equiv u_j^2$. The reason was traced back to the presence of breathers
(that arise as a result of a modulational instability \cite{RumpfN,rassegne})
and attributed to the weakness of their coupling to the background. 

\begin{figure}
\includegraphics[clip, width=7cm]{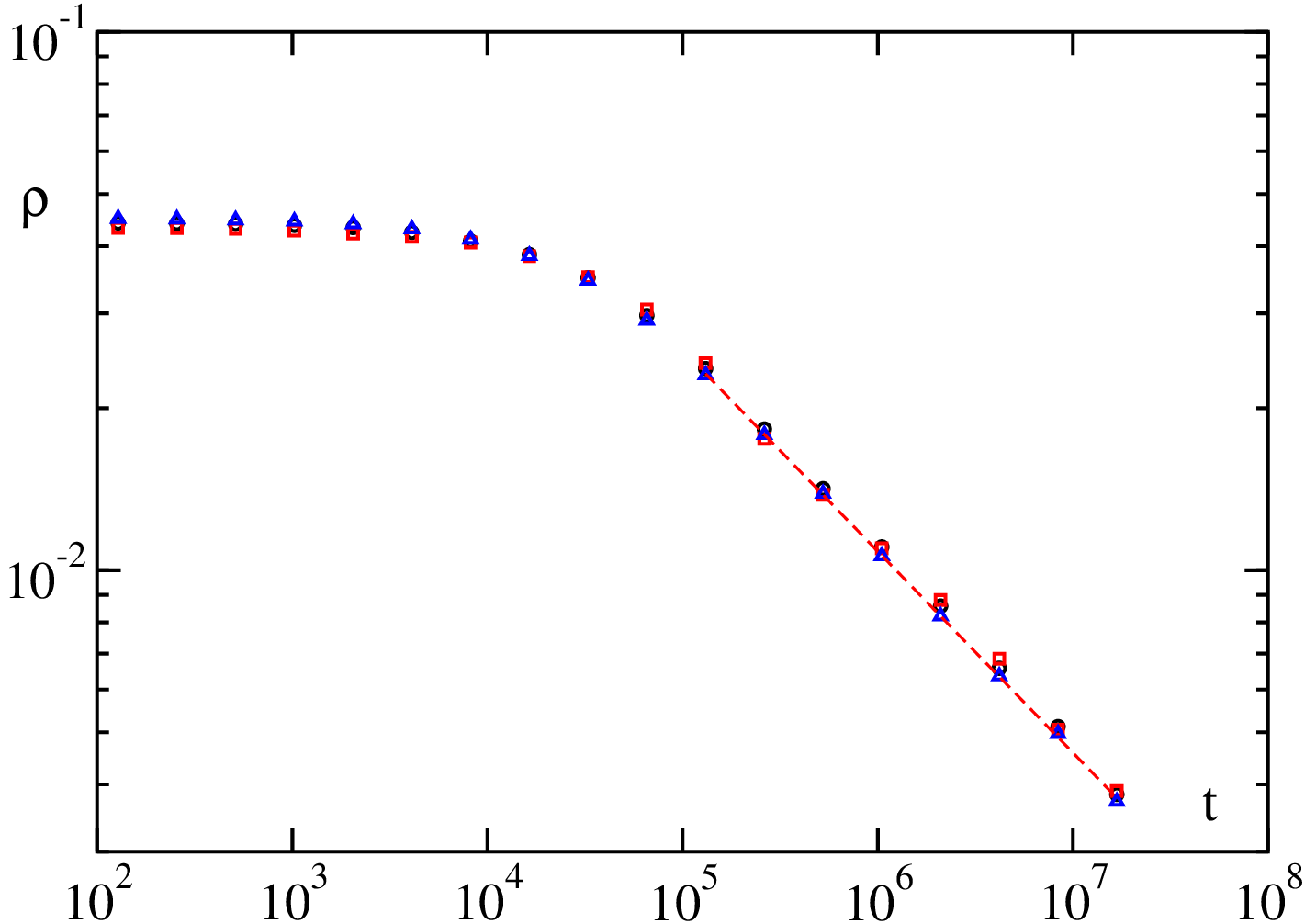}
\caption{(Color online). Average breather density $\rho$ versus time $t$
in a DNLS chain as from
MMC simulations in the weak-coupling limit. Circles, squares and triangles
refer to $N=400$, 1600, and 6400, respectively. The straight line with slope
0.37 is the result of a power-law fit.}
\label{MC}
\end{figure}

\begin{figure}
\includegraphics[clip, width=7cm]{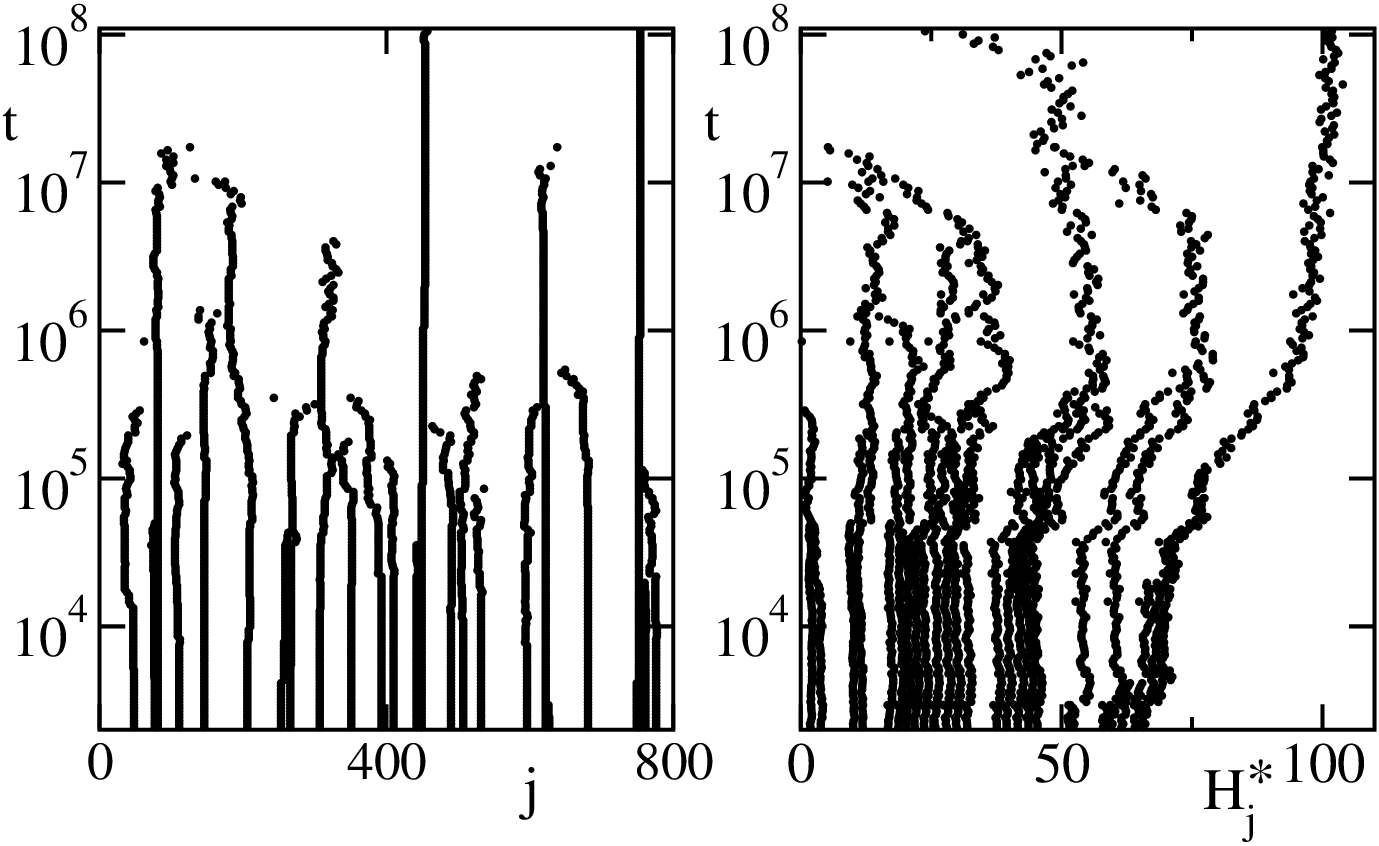}
\caption{(a) Space time representation of MMC dynamics of breathers in the NT
region. (b) Same as in panel a, with the label $j$ replaced by $H^*_j$, see
text.}
\label{MC2}
\end{figure}

{\it Microcanonical Monte Carlo simulations.} 
We start our analysis of DNLS dynamics from the weak-coupling limit
(i.e. large particle densities). In this regime, the region $\mathrm{R_n}$
corresponds to $h/a^2>2$ \cite{rumpf} (with our normalizations). The motivation
for a MMC study is a direct investigation of entropic effects during the time
evolution. In order to mimic the deterministic evolution, we have
introduced a local updating rule which applies to a randomly selected triplet
of consecutive sites. Since it is necessary to leave the number of particles
$A_j^s = A_{j-1} + A_j + A_{j+1}$ and the energy
$H_j^s = A_{j-1}^2 + A_j^2 + A_{j+1}^2$ unchanged, 
the new configuration must lie along the
intersection between a plane and the surface of a sphere, while restricted to
the positive octant ($A_j\ge 0$). Depending on the relative amplitude of
$H_j^s$ and $A_j^s$,
this line may be either a full circle or made of three separate arcs.
In the former case, we select a point on the circle (assuming uniform
probability); in the latter one, we select a point within the same arc as of
the IC. It is easy to verify that detailed balance is satisfied.
After having observed that the above algorithm yields a fast convergence
towards the expected equilibrium state in the $T>0$ region, we have performed
some simulations in $\mathrm{R_n}$. The IC is fixed by assigning
a random amplitude $A_j$ (with a flat distribution between 0 and 1) to all
sites and adding 40/3 units to 5\% of sites randomly chosen over the chain. 
This choice implies that $h/a^2\approx 7>2$, i.e. the IC lies inside
$\mathrm{R_n}$. In Fig.~\ref{MC} we plot the MMC evolution of the fraction
$\rho$ of sites, where the amplitude $A_j$ is larger than the threshold
$\eta = 10$ \cite{nnote} for three different system sizes \cite{foot}.
The three curves nicely overlap and show for large time $t$ a clear power-law decay
$\rho \sim t^{-\alpha}$, with an exponent $\alpha \approx 0.37$. Time $t$ is
given by the number of MMC steps divided by $N$. This
means that the system converges, although slowly, to the equilibrium
state characterized by a single breather (i.e. $\rho=0$ in the thermodynamic
limit).
In order to gain further insight, we have looked at the MMC
evolution in two different ways. In Fig.~\ref{MC2}a, we plot the position of the
breathers of amplitude larger than $\eta$. The breathers are basically static
and either terminate abruptly, or eventually exhibit a weak mutual attraction.
The representation in panel Fig.~\ref{MC2}b is more
illuminating. There, we plot the sum of the ``excess" energies
$H^*_j = \sum_{l=1}^j (A_l^2- \eta^2) \theta(A_l^2- \eta^2)$ ($\theta$
is the Heaviside function)
in correspondence of each breather. The abscissa of the rightmost breather
is the total excess energy contained in the breathers. After an initial
transient during which the excess energy is constant, the breathers perform a
sort of Brownian motion in energy space until they collide and merge. 
This (sub)diffusive motion reveals a slow convergence that is due to a
a likewise coarsening phenomenon. It is intriguing to see that the value of
$\alpha$ (0.37) is relatively close to the critical exponent of the
Cahn-Hilliard equation ($1/3$). We leave the task of identifying the correct
universality class to future investigations.

\begin{figure}
\includegraphics[clip, width=7cm]{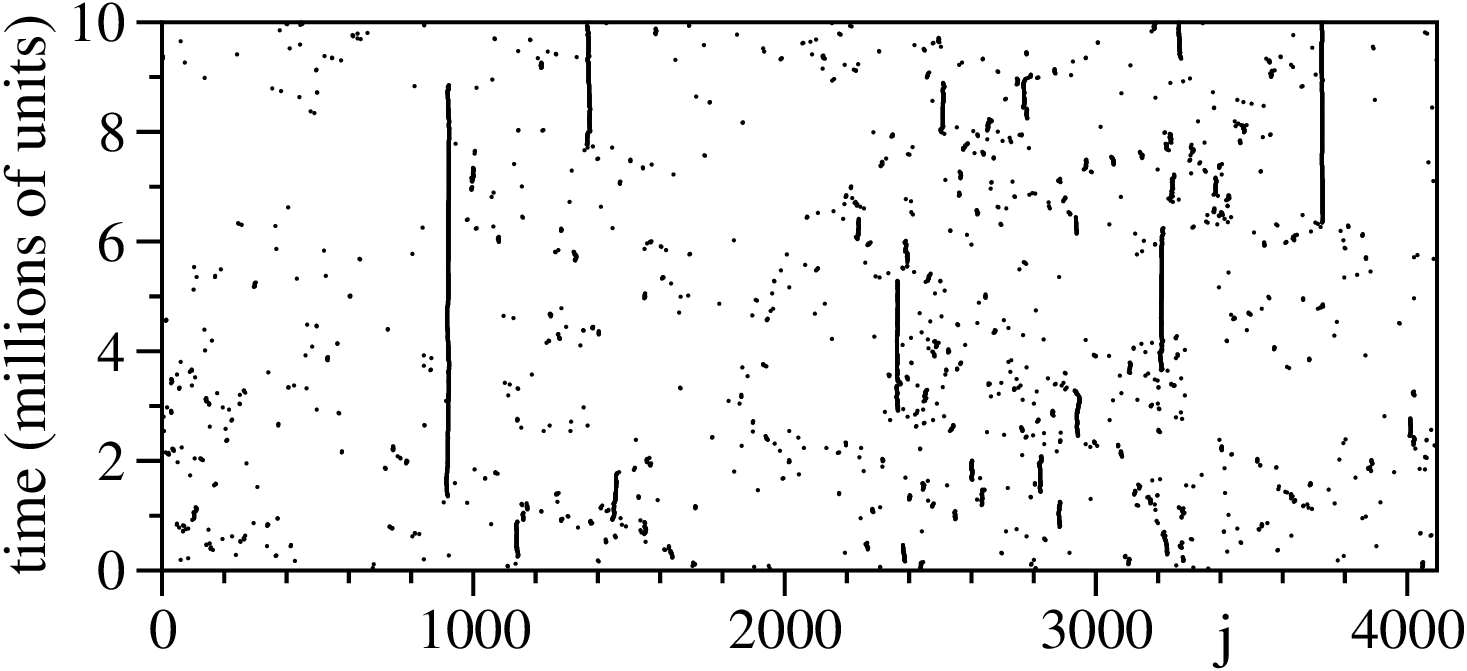}
\caption{Evolution of the local amplitude in an NT states (dots correspond to
points where $A_j>10$) for $a=1$ and $h=2.4$.}
\label{pattern}
\end{figure}

\begin{figure}
\includegraphics[clip, width=7cm]{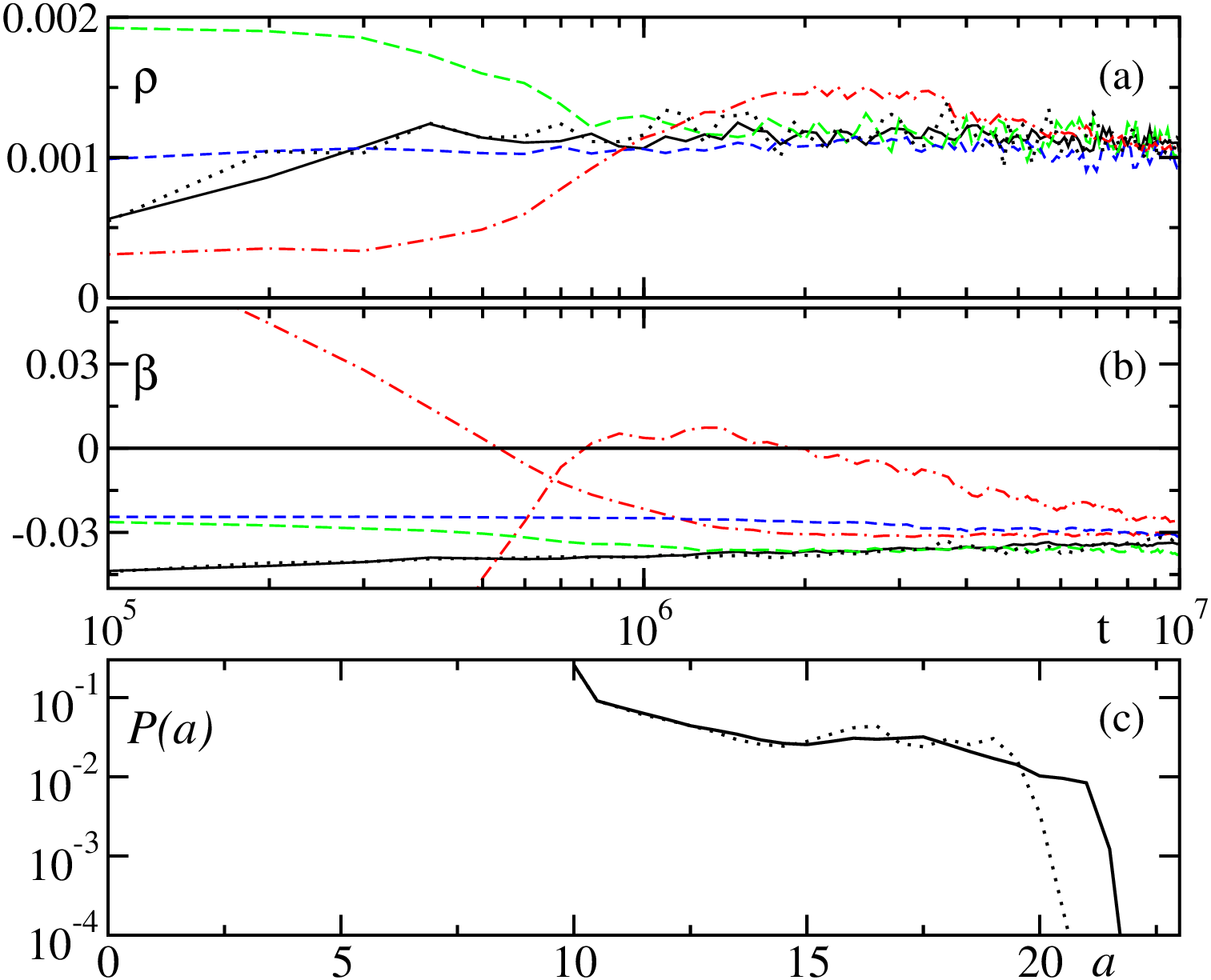}
\caption{(Color online) Evolution for $a=1$, $h=2.4$ of (a) the breather density $\rho$
and (b) the inverse temperature $\beta = 1/T$ starting from
different ICs for a chain with $N=4096$: solid, long--dashed and 
dashed lines refer to a suitable homogeneous background plus 0, 
4 (amplitude 20) and 8 (amplitude 16) 
breathers, respectively, while the dot-dashed curves refer to half a chain
empty. Both $\rho$ and $\beta$
are averaged over a moving window of $10^5$ time units. Notice that 
in panel (b) the two dot--dashed lines correspond to 
spatial averages in the initially empty and filled sectors of the chain.
(c) Breather probability density after a transient of $5 \cdot 10^{6}$
time units. Data are not collected below the threshold $\eta=10$.
In the three panels the dotted lines refer to homogeneous ICs and $N=2048$.
All data sets have been also averaged over a dozen different realizations
of each IC.
}
\label{running}
\end{figure}

{\it DNLS Dynamics.}
The MMC study has shown that in the weak-coupling limit, the convergence
is characterized by a coarsening process: the fewer the breathers, the weaker
their interaction and the slower the convergence process. The uncoupled limit 
is, however, rather peculiar, as it lacks the hopping mechanism. In the natural
case of a finite coupling, we have turned our attention to simulations of the 
deterministic DNLS since effective implementations of MMC rules are not feasible.
Fig.~\ref{pattern} presents the space-time evolution for a chain of $4096$ sites
for  $a=1$ and $h=2.4$. To emphasize the breather dynamics, we plot only the
lattice points, where the instantaneous mass $A_j$ is larger than $\eta=10$. 
Thanks to a symplectic 4th-order algorithm of Yoshida type \cite{xx}, we have
simulated the dynamics over time scales much longer than in the previous
numerical studies. In Fig.~\ref{pattern} one can see that, like in the MMC
simulations, the breathers do not basically move; at variance with the uncoupled 
case, however, we observe spontaneous birth and death of breathers as in a 
standard stationary process.
This is due to the presence of a finite interaction (hopping) energy: the
background can store excess energy in the phase differences of neighbouring
sites and this implies that breathers can spontaneously nucleate.
In order to better investigate the convergence properties of the dynamics, we
have built four different sets of ICs, all with the same energy and number of
particles but substantially different macroscopic structures (see the caption
of Fig.~\ref{running}). First we have monitored the evolution of the
density of breathers (identified by setting $\eta=10$). The resulting average
results for $N=4096$ are plotted in Fig.~\ref{running}a. At variance with the
simulations described in Fig.~\ref{MC}, the density appears to convergence
towards a common finite value, i.e. towards a multi-breather stationary state.
Moreover, the dotted line corresponding to a simulation with $N=2048$
indicates that the asymptotic value of $\rho$ is independent of the system size
(and approximately equal to one breather per thousand sites).
We have also monitored the temperature. In this model, the
standard kinetic definition does not apply, since the Hamiltonian is not
decomposable into kinetic and potential energy. Nevertheless, one can use the
microcanonical definition, $T^{-1}=\partial{S}/\partial{H}$, where $S$ is the
thermodynamic entropy and the partial derivative must be computed taking into
account the existence of two conserved quantities. This task requires long and
careful calculations with a final {\it non-local} expression \cite{franzo}. 
We have preliminarily verified that the definition of \cite{franzo} works for 
positive temperatures when applied to the full and to short sub chains. The 
results in $\mathrm{R_n}$ are plotted in Fig.~\ref{running}b, where
one can see that $\beta\equiv 1/T$ converges to a common value for all of the
four different classes of ICs (either from above or from below), with
negligible finite-size effects (once again the dotted curve refers to a chain
of half length, $N=2048$). Altogether, one can conclude that on the time scales
that are numerically accessible (of order $10^7$), the DNLS converges towards a
multi-breather state characterized by a finite density of breathers and a
well-defined negative temperature.

In practice, what happens is that the breathers are prevented from becoming
``too large". In Fig.~\ref{running}c, one can indeed see that the
pseudo-stationary distribution of breather amplitudes is effectively localized
below $a= 22$ and, more importantly, the distribution is basically independent of
the system size. Altogether, this means that there is a dynamical process 
which ``screens" the high amplitude region. One can argue that this dynamical 
effect is due to the low efficiency of the energy--transfer among breathers 
interacting through the background, a manifestation of their intrinsic dynamical 
stability. As a consequence, the evolution is ``confined" to a region of the 
phase space that is characterized by a NT. We still expect a convergence towards 
the equilibrium state predicted in \cite{rumpf} eventually, but the process occurs 
on such long (unobservable) time scales \cite{nota} (further slowed down by coarsening
processes) to make it unaccessible. It is nevertheless worth mentioning
that if the IC contains one or more sufficiently-high breathers, the amount of
energy that they carry is effectively frozen, while the structure of the
metastable state depends on the remaining ``excess" energy that is
still free to diffuse between background and smaller breathers. Altogether,
the scenario is quite reminiscent of what happens in spin glasses although a 
more quantitative analysis in necessary to clarify for example possible aging
phenomena.

{\it Generation of NT states.} 
NT states can be generated by moving in parameter space from an initial state 
in a region where equilibrium is characterized by a homogeneous regime, i.e.
positive temperatures. Two particularly simple mechanisms to generate NT states are: 
localized dissipations with removal of mass (and energy) at  the extrema of the
chain \cite{LFO} or a free expansion in lattices of larger size. In the inset of 
Fig.~\ref{paramsp} we show an average trajectory that results from the
first method of boundary dissipation in a chain with $N=200$ sites. The free 
expansion method from a positive-temperature equilibrium state to a NT sate is 
demonstrated in the upper dot-dashed line in Fig.~\ref{running}b). From Fig.~\ref{paramsp}, 
one can easily realize that a free expansion amounts to decreasing $a$ until the
$T=\infty$ line is crossed and the NT region is eventually reached. Thus, in either
BECs in optical lattices, or arrays of optical waveguides, this
procedure amounts to prepare a standard equilibrium state at $T>0$ and leave it
to spread in sufficiently larger lattice structures. This method is much simpler
than the experimental schemes described in \cite{mosk05}.

{\it Conclusions.}
We have explored the dynamical behavior of the DNLS equation in the parameter 
region corresponding to NT states. Numerical simulations show that for sufficiently 
uniform initial conditions, the system converges towards a stationary although 
metastable state characterized by a negative temperature and a finite density of 
breathers. The asymptotic convergence to the single-breather equilibrium state 
occurs on much longer (and practically unattainable) time scales. We envisage a 
quantitative theory that relates the time scales to the breather amplitudes, their 
stationary density and the system size so as to establish possible connections with, 
e.g. aging phenomena.

{\it Acknowledgements.} RL acknowledges financial support from the Italian 
MIUR-PRIN project n. 20083C8XFZ and G-LO from the EU Commission FET Open 
Grant HIDEAS. SI and AP thank T. Carletti for useful discussions 
on the implementation of symplectic integrators.


\begin{thebibliography}{99}
\bibitem{nnote} When we refer to the local amplitude we always refer to the
square amplitude.
\bibitem{foot} We have selected a sufficiently large threshold to ensure a
negligible contribution of the background fluctuations, here expected to
follow a Poisson distribution \cite{rumpf}.
\bibitem{nota} In \cite{rasm1}  it was already noticed that a single breather state
was evolving so slowly as to converge towards the equilibrium state over
$10^{70}$ time units.
\end{thebibliography}
\end{document}